\documentstyle[preprint,eqsecnum,aps,psfig]{revtex}
\begin{document}
\draft
\title{$\Xi$ and $\overline{\Xi}$ Production in 158 GeV/Nucleon Pb+Pb Collisions}
\author{(The NA49 Collaboration)\\
H.~Appelsh\"{a}user$^{7}$~\cite{sb}, J.~B\"{a}chler$^{5}$,
S.J.~Bailey$^{17}$, D.~Barna$^{4}$, L.S.~Barnby$^{3}$,
J.~Bartke$^{6}$, R.A.~Barton$^{3}$, H.~Bia{\l}\-kowska$^{15}$,
A.~Billmeier$^{10}$, C.O.~Blyth$^{3}$, R.~Bock$^{7}$,
C.~Bormann$^{10}$, F.P.~Brady$^{8}$, R.~Brockmann$^{7}$~\cite{th},
R.~Brun$^{5}$, P.~Bun\v{c}i\'{c}$^{5,10}$, H.L.~Caines$^{3}$,
L.D.~Carr$^{17}$, D.~Cebra$^{8}$, G.E.~Cooper$^{2}$,
J.G.~Cramer$^{17}$, M.~Cristinziani$^{13}$, P.~Csato$^{4}$,
J.~Dunn$^{8}$, V.~Eckardt$^{14}$, F.~Eckhardt$^{13}$,
M.I.~Ferguson$^{5}$, H.G.~Fischer$^{5}$, D.~Flierl$^{10}$,
Z.~Fodor$^{4}$, P.~Foka$^{10}$, P.~Freund$^{14}$, V.~Friese$^{13}$,
M.~Fuchs$^{10}$, F.~Gabler$^{10}$, R.~Ganz$^{14}$, W.~Geist$^{14}$, J.~Gal$^{4}$,
M.~Ga\'zdzicki$^{10}$, E.~G{\l}adysz$^{6}$, J.~Grebieszkow$^{16}$,
J.~G\"{u}nther$^{10}$, J.W.~Harris$^{18}$, S.~Hegyi$^{4}$,
T.~Henkel$^{13}$, L.A.~Hill$^{3}$, H.~H\"{u}mmler$^{10}$~\cite{als},
G.~Igo$^{12}$, D.~Irmscher$^{7}$, P.~Jacobs$^{2}$, P.G.~Jones$^{3}$,
K.~Kadija$^{19,14}$, V.I.~Kolesnikov$^{9}$, A.~Konashenok$^{2}$, M.~Kowalski$^{6}$,
B.~Lasiuk$^{12,18}$, P.~L\'{e}vai$^{4}$, F.~Liu$^{10}$, A.I.~Malakhov$^{9}$,
S.~Margetis$^{11}$, C.~Markert$^{7}$, G.L.~Melkumov$^{9}$,
A.~Mock$^{14}$, J.~Moln\'{a}r$^{4}$, J.M.~Nelson$^{3}$,
M.~Oldenburg$^{10}$, G.~Odyniec$^{2}$, G.~Palla$^{4}$,
A.D.~Panagiotou$^{1}$, A.~Petridis$^{1}$, A.~Piper$^{13}$,
R.J.~Porter$^{2}$, A.M.~Poskanzer$^{2}$, D.J.~Prindle$^{17}$,
F.~P\"{u}hlhofer$^{13}$, W.~Rauch$^{14}$, J.G.~Reid$^{17}$,
R.~Renfordt$^{10}$, W.~Retyk$^{16}$, H.G.~Ritter$^{2}$,
D.~R\"{o}hrich$^{10}$, C.~Roland$^{7}$, G.~Roland$^{10}$,
H.~Rudolph$^{10}$, A.~Rybicki$^{6}$, A.~Sandoval$^{7}$, H.~Sann$^{7}$,
A.Yu.~Semenov$^{9}$, E.~Sch\"{a}fer$^{14}$, D.~Schmischke$^{10}$,
N.~Schmitz$^{14}$, S.~Sch\"{o}nfelder$^{14}$, P.~Seyboth$^{14}$,
J.~Seyerlein$^{14}$, F.~Sikler$^{4}$, E.~Skrzypczak$^{16}$,
R.~Snellings$^{2}$, G.T.A.~Squier$^{3}$, R.~Stock$^{10}$,
H.~Str\"{o}bele$^{10}$, Chr.~Struck$^{13}$, T.~Susa$^{19}$, I.~Szentpetery$^{4}$,
J.~Sziklai$^{4}$, M.~Toy$^{2,12}$, T.A.~Trainor$^{17}$,
S.~Trentalange$^{12}$, T.~Ullrich$^{18}$, M.~Vassiliou$^{1}$,
G.~Veres$^{4}$, G.~Vesztergombi$^{4}$, D.~Vrani\'{c}$^{5,19}$,
F.~Wang$^{2}$, D.D.~Weerasundara$^{17}$, S.~Wenig$^{5}$,
C.~Whitten$^{12}$, T.~Wienold$^{2}$~\cite{sb}, L.~Wood$^{8}$, N.~Xu$^{2}$,
T.A.~Yates$^{3}$, J.~Zimanyi$^{4}$, X.-Z.~Zhu$^{17}$, R.~Zybert$^{3}$}
\address{$^{1}$Department of Physics, University of Athens, Athens, Greece.\\
$^{2}$Lawrence Berkeley National Laboratory, University of California, Berkeley, USA.\\
$^{3}$Birmingham University, Birmingham, England.\\
$^{4}$KFKI Research Institute for Particle and Nuclear Physics, Budapest, Hungary.\\
$^{5}$CERN, Geneva, Switzerland.\\
$^{6}$Institute of Nuclear Physics, Cracow, Poland.\\
$^{7}$Gesellschaft f\"{u}r Schwerionenforschung (GSI), Darmstadt, Germany.\\
$^{8}$University of California at Davis, Davis, USA.\\
$^{9}$Joint Institute for Nuclear Research, Dubna, Russia.\\
$^{10}$Fachbereich Physik der Universit\"{a}t, Frankfurt, Germany.\\
$^{11}$Kent State University, Kent, OH, USA.\\
$^{12}$University of California at Los Angeles, Los Angeles, USA.\\
$^{13}$Fachbereich Physik der Universit\"{a}t, Marburg, Germany.\\
$^{14}$Max-Planck-Institut f\"{u}r Physik, Munich, Germany.\\
$^{15}$Institute for Nuclear Studies, Warsaw, Poland.\\
$^{16}$Institute for Experimental Physics, University of Warsaw, Warsaw, Poland.\\
$^{17}$Nuclear Physics Laboratory, University of Washington, Seattle, WA, USA.\\
$^{18}$Yale University, New Haven, CT, USA.\\
$^{19}$Rudjer Boskovic Institute, Zagreb, Croatia.\\}
%
%
\date{\today}
\maketitle
\begin{abstract}
%
We report measurements of $\Xi^-$ and $\overline{\Xi}^+$ hyperon absolute yields as a function of rapidity in 158 GeV/c Pb+Pb collisions. At midrapidity, $\frac{dN}{dy} =
2.29 \pm 0.12$ for $\Xi^-$, and $0.52 \pm 0.05$ for $\overline{\Xi}^+$, leading to the ratio of $\overline{\Xi}^+$/$\Xi^-$=0.23$\pm$0.03. Inverse slope
parameters fitted to the measured transverse mass spectra are of the order of $T \approx 300$ MeV near mid-rapidity. The estimated total yield of $\Xi^-$ particles in Pb+Pb central interactions amounts to 7.4$\pm$1.0 per collision. Comparison to $\Xi^-$ p
roduction in properly scaled p+p reactions at the same energy reveals a dramatic enhancement (about one order of magnitude) of $\Xi^-$ production in Pb+Pb central collisions over elementary hadron interactions.
\end{abstract}
%
%
Ultra-relativistic nuclear collisions provide a unique opportunity to create and study a hypothetical new phase of nuclear matter: the quark-gluon plasma (QGP). For an overview see \cite{QM97} and references therein. In the color-deconfined QGP phase, a g
eneral enhancement of strange-sector particle yields with respect to that from hadronic interactions, has been predicted \cite{KMR,Raf82}. This stems from the argument that the high parton density and lower energy threshold for $s\bar{s}$ quark-pair produ
ction in the plasma, compared to a hadron gas, may lead to far shorter equilibration times in the plasma than in a hadron gas. Thus, a high characteristic strangeness yield would result from a plasma phase essentially unaltered during the subsequent hadro
nic expansion. Moreover, in an environment with high baryon density, as e.g. in Pb+Pb collisions at the SPS \cite{Rol}, $s\bar{s}$ pair formation should be favored if the lowest available $u$- and $d$-quark energy levels are above 2$m_s$.

The strangeness enhancement and its relation to QGP formation have been studied within the QCD framework \cite{KMR1,Com}, confirming earlier predictions from kinetic equilibrium models \cite{KMR,Raf82}.
In lowest-order perturbative QCD, $s\bar{s}$ quark pairs are created
via gluonic fusion ($gg \rightarrow s\bar{s}$) and light
quark-antiquark annihilation ($q\bar{q}\rightarrow s\bar{s}$).
In a kinetically equilibrated partonic gas at sufficient temperature
($T \approx m_s$), the gluonic production dominates ($\sim$90\%), leading to
chemical equilibration times comparable to the expected plasma lifetime
\cite{KMR1}. Relaxation times for chemical equilibrium via
$q\bar{q}$ annihilation are too slow \cite{Biro}.
The gluonic degrees of freedom, transiently present during collisions, would then
result in a high abundance of strangeness. Studies of strangeness production in central A+A collisions may thus help to ascertain the existence of a transient partonic phase.

Multi-strange hyperon production is expected to be
particularly sensitive to the rapid flavor equilibration in the early
stages of the collisions, when the energy density is highest and the
phase transition is most likely \cite{Raf91}. Multi-strange baryons and antibaryons should coalesce easily during
hadronization of the QGP and should survive subsequent hadronic interactions
since their inelastic cross sections are generally small.
Thus, a QGP transition should significantly enhance multi-strange
hyperon yields. In fact, a recent publication of $\Lambda$, $\Xi$, and $\Omega$ particle yields near mid-rapidity by the WA97 Collaboration \cite{And} has demonstrated such an enhancement over p+Pb collisions. In a complementary study we obtained estimate
s of the total 4$\pi$ yields of $\Xi^-$ and $\overline{\Xi}^+$. We shall derive an enhancement of about a factor 10 of the $\Xi^-$ particle yields in central Pb+Pb collisions over and above properly scaled pp collision yields at similar energy.

Studies of $\Xi^-$ and $\overline{\Xi}^+$ (S=$\pm$2) production in Pb+Pb collisions at 158 GeV/c at the CERN SPS were carried out by the NA49 collaboration using two independent analysis techniques.
The NA49 experiment uses four large Time Projection Chambers (TPC's) for tracking and momentum analysis. Two of them, Vertex TPC-1 and -2 (VTPC1,2) are placed inside a magnetic field (with a total bending power of 9~T$\cdot$m), and two Main TPCs (MTPC's) 
are placed further downstream outside the magnetic field on either side of the beam. A detailed description of the apparatus can be found in \cite{Jon}. In addition to the standard configuration (target in front of VTPC1) a set of central Pb+Pb events was
 taken with the target placed about 20 cm upstream of the MTPC entrance windows. $\Xi$ particles were studied in the VTPC2 detector (traditional approach, using a magnetic field and direct momentum measurements, target in front of the VTPC1) and in the MT
PC detector (special configuration, target in front of the MTPC, no magnetic field).
The lack of charge information in MTPC runs (special configuration) caused by the absence of a magnetic field did not allow a separation between $\Xi^-$ and $\overline{\Xi}^+$ to be made. Thus the sum of $\Xi^-$+$\overline{\Xi}^+$ was measured with the MT
PCs.
The cascade decay
\begin{displaymath}
\Xi^-\rightarrow\Lambda +\pi^- ~~~({\rm where}~~
\Lambda \rightarrow p+\pi^-)
\end{displaymath}
and the corresponding $\overline{\Xi}^+$ decay have a characteristic topology: the first decay vertex has the appearence of a broken line (``kink'') followed by the V$^0$ ($\Lambda$) decay. A V$^0$ points back to a ``kink'' rather than to the primary even
t vertex. $\Lambda$ particles were identified by reconstructing their decays into a final state containing only charged particles.

Measurements of $\Xi$ particles in heavy ion collisions in the presence of a magnetic field have been reported previously (WA85/94/97\cite{Kra}, NA35\cite{Ret}, and NA36\cite{NA36} experiments) and the methodology is well established.
For completeness, we briefly outline the two main steps of this analysis:
\begin{itemize}
\item
reconstruction and identification of $\Lambda$s and $\overline{\Lambda}$s is done by measuring their charged decay products (V$^0$ decay). Momenta and charges of the decay products are obtained from the curvature of the trajectories, whereas momentum and 
mass of the parent are derived from energy/momentum conservation assuming the decay scheme, e.g. $\Lambda \rightarrow p+\pi ^-$.
\item
the second cascade decay product (so-called ``$\pi$-batchelor'' arising from the ``kink'' vertex) is found for each $\Lambda$ ($\overline{\Lambda}$) candidate by combining all positive (negative) tracks which did not point back to the target, with the rec
onstructed $\Lambda$ ($\overline{\Lambda}$) in order to determine whether they could originate from a common vertex. For successful pairs the $\Xi^-$ ($\overline{\Xi}^+$)  momentum is derived using momentum/energy conservation. The $\Xi^-$ and $\overline{
\Xi}^+$ candidates are required to point back to the primary vertex. 
\end{itemize}

\noindent
Our analysis of Vertex-TPC data follows this approach. 

The idea of detecting $\Xi$s in tracking detectors without the presence of a magnetic field is new, and therefore requires more explanation.
with all verticies visible vs. TPC with no direct information on verticies) and significantly lighter environment (track multiplicity lower by one order of magnitude) - the methods developed by the UA5 collaboration were not applicable to the analysis of 
heavy ion interactions. 
\noindent
A fairly detailed description of this method can be found in \cite{me}; here we only outline the main steps:

\begin{itemize}
\item
reconstruction of V$^0$-type vertices done the conventional way (by combining track pairs), but with straight, not curved, tracks \cite{Mar}.
\item
for each $\Lambda$ candidate a search is performed to find the 
matching ``$\pi$-batchelor''. The match criterion was coplanarity of the $\pi$-batchelor, the $\Lambda$ decay vertex and the interaction vertex. 
\item
roughly coplanar candidates together with the preliminary estimates of both vertex positions are fed into a geometrical fit routine \cite{CERN}.
The coplanarity constraint used in the fitting routines turned out to be very efficient in eliminating the majority of the combinatorial background.
\item
momenta of parent and daughter particles are reconstructed in each vertex from the decay angles (while their masses are assumed). This is done by requiring energy/momentum conservation at each vertex, separately.
\item
using the $\Lambda$ momentum, calculated at the V$^0$ vertex, in 
the energy/momentum conservation equations for the ``kink'' vertex, allowed the $\Xi$ mass to be calculated, rather than assumed.
\end{itemize}

The advantages of this approach include the easier/higher-accuracy (straight-line) tracking in the large MTPC volumes, the absence of so called ``E$\otimes$B distortions'' resulting from an inhomogeneus magnetic field, and the proximity of the target to t
he active volume. The final results are limited, however, to the combined ($\Xi^-$ + $\overline{\Xi}^+$) yields.

\begin{figure}[b]
\centerline{\psfig{figure=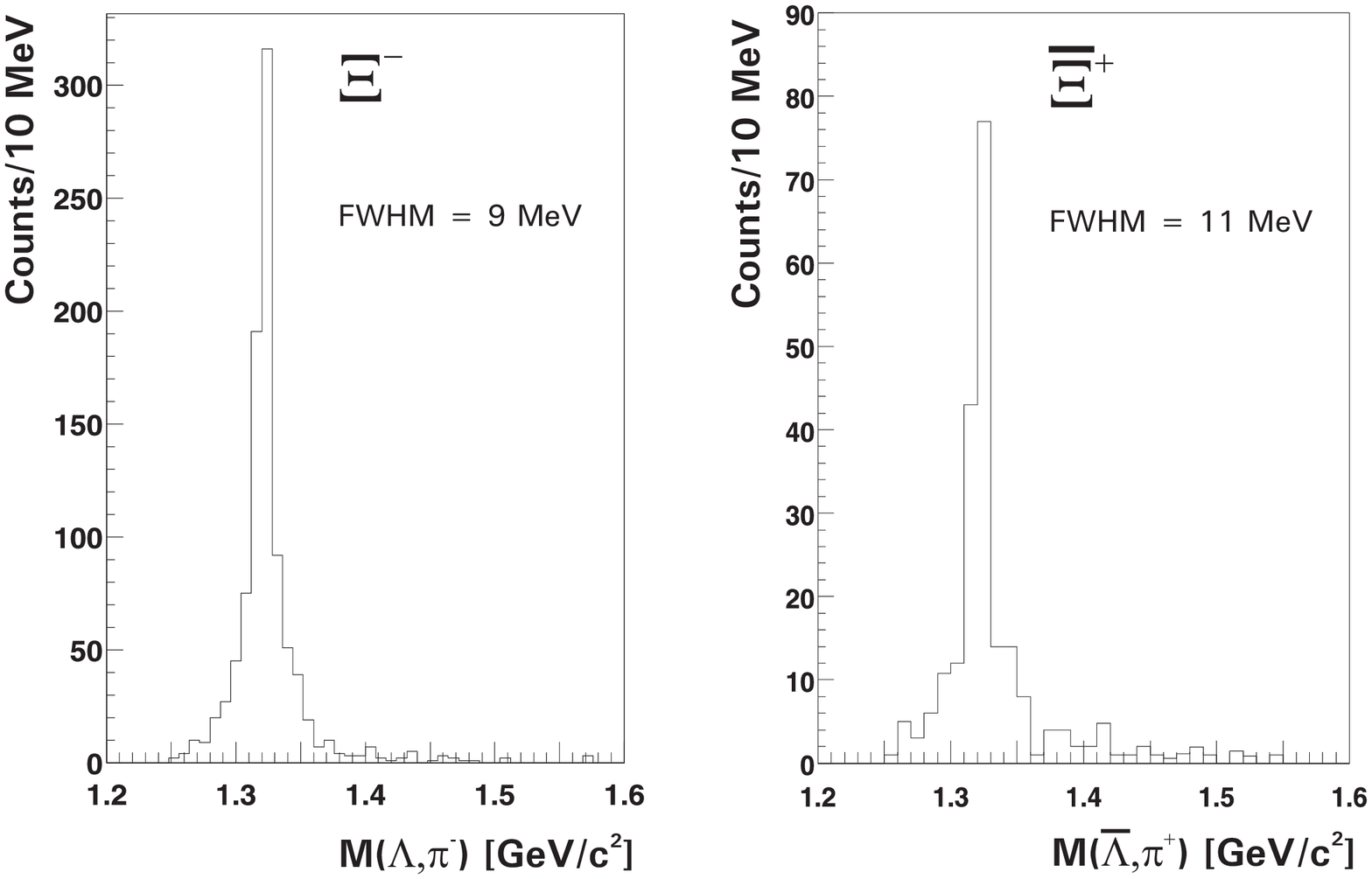,height=5.2in,width=7.4in}}
\caption{
Invariant mass spectra for $\Xi^-$ and $\overline{\Xi}^+$.
}
\label{Fig1}
\end{figure}

\begin{figure}[b]
\centerline{\psfig{figure=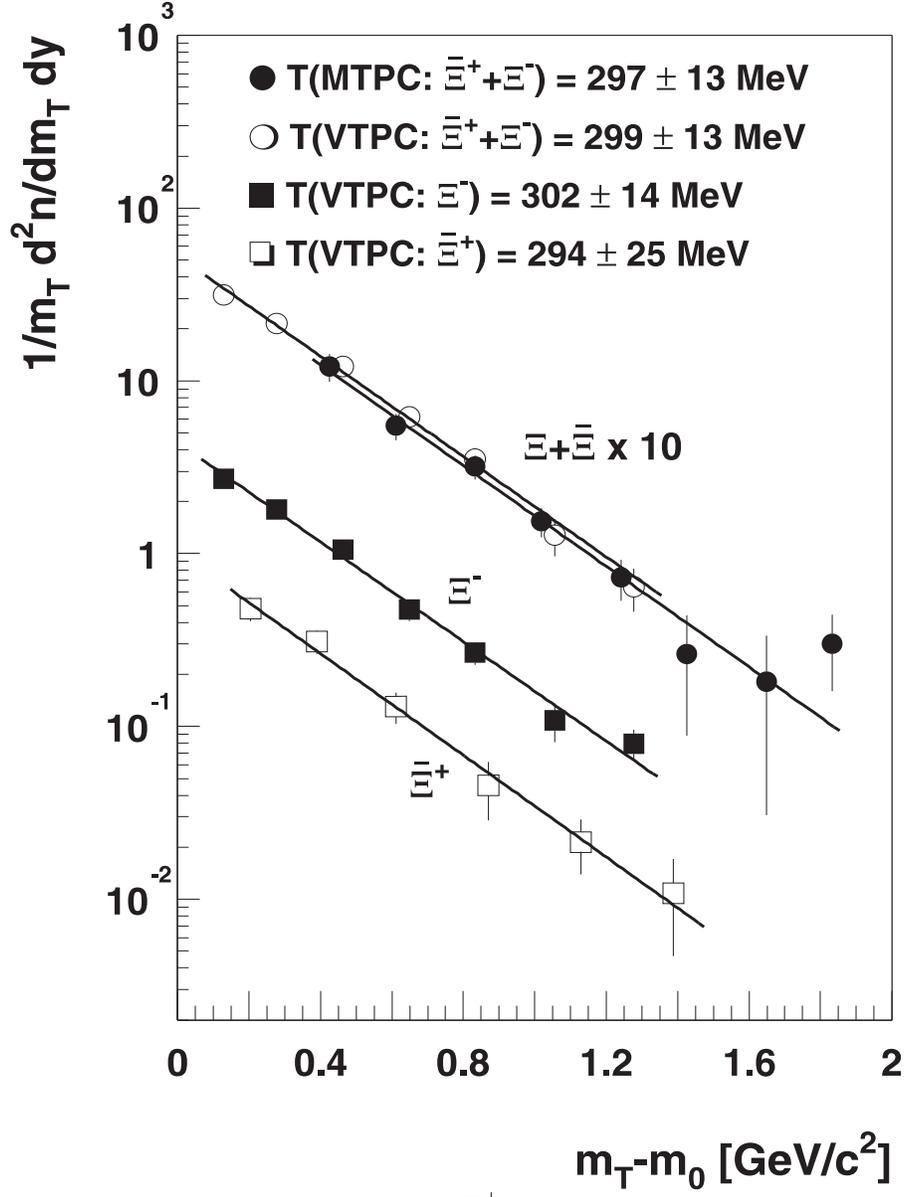,height=6.22in,width=4.66in}}
\caption{
Transverse mass spectra for $\Xi^-$ and $\overline{\Xi}^+$ analyzed in VTPC2, and $\Xi^-$ + $\overline{\Xi}^+$ analyzed in MTPC (offset by factor 10)
}
\label{Fig2}
\end{figure}

\begin{figure}[b]
\centerline{\psfig{figure=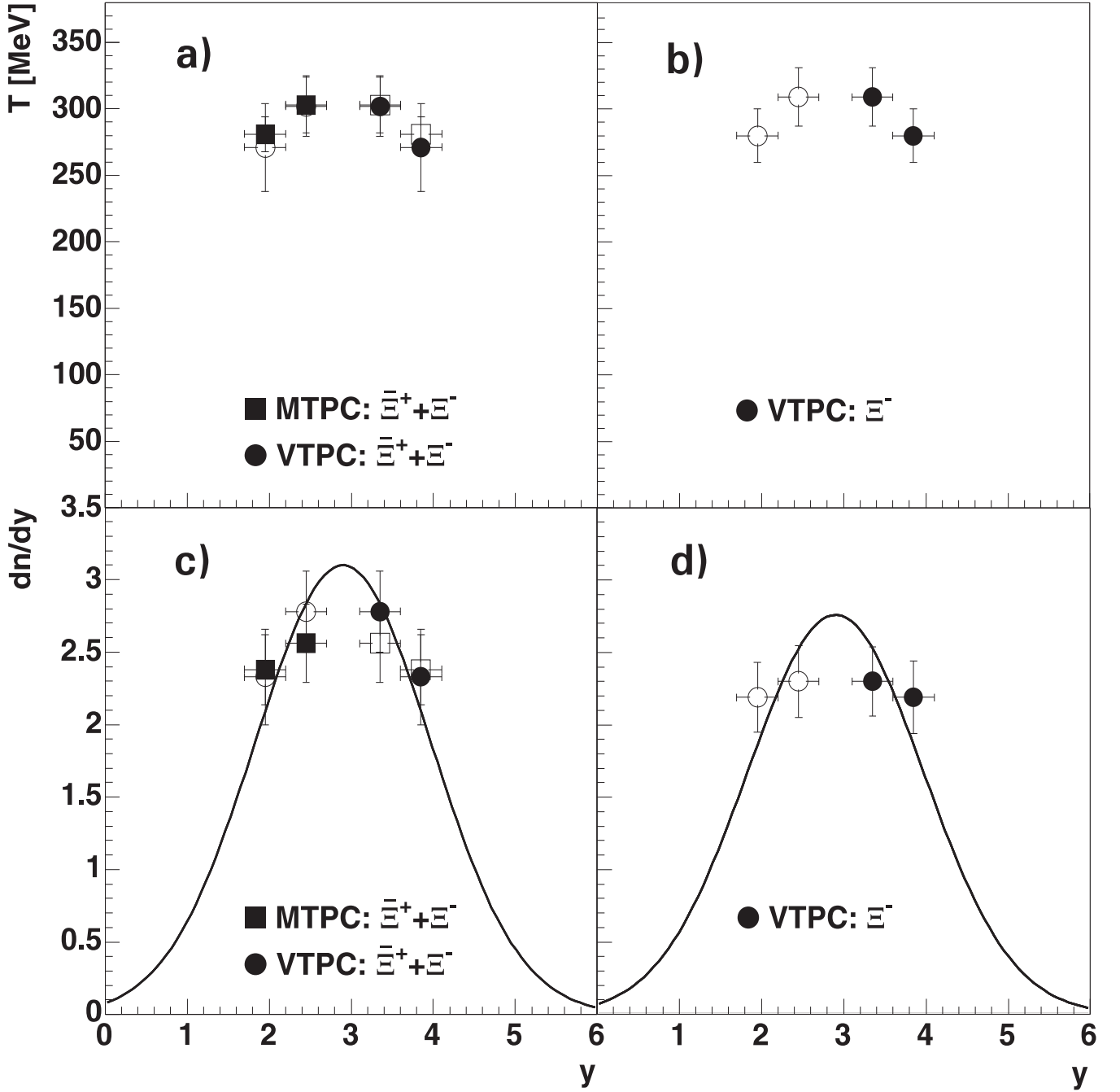,height=7.0in,width=7.0in}}
\caption{
a,b:Inverse slope parameters of $\Xi^-$ + $\overline{\Xi}^+$ and $\Xi^-$ (right) as a function of $\Xi$ rapidity.
c,d:Rapidity distributions of $\Xi^-$ + $\overline{\Xi}^+$ and $\Xi^-$ (right). The open symbols are the measured points reflected at y$_{mid}$=2.9. See text for details.
}
\label{Fig3}
\end{figure}

The data presented here were obtained during two different runs, both with a central trigger. The trigger selected the most central 5~\% of the total inelastic cross section ($\sim$7~barn) for the standard configuration (analysis in the VTPC2), and 7~\% -
 for the special configuration (analysis in MTPC). It corresponds to maximum impact parameters of $b$=3.5 and 4.0~fm, respectively, or an average number of 370 participants \cite{Alb}. The phase-space coverage of the two detectors is chosen symmetrically 
on either side of mid-rapidity ($y_{mid}$=2.9): the MTPC covers $y\in$(1.7-2.7) and $p_T\geq$0.9 GeV/c (backward hemisphere), whereas the VTPC2 covers forward hemisphere: $y\in$(3.1-4.1) and $p_T\geq$0.5 GeV/c. Both measurements together cover the bulk of
 the $\Xi$ particle yield. The data sample consists of 58K events from the VTPC2, and 240K events from the MTPC. Analysis of the data taken with the magnetic field resulted in 720 $\Xi^-$ and 138 $\overline{\Xi}^+$ candidates, whereas analysis of the data
 taken without magnetic field gave 2000 $\Xi^-$ + $\overline{\Xi}^+$ candidates. Various cuts were applied to remove the combinatorial background; most of them being geometrical. For details see \cite{Gab}. Besides the coplanarity requirement used in the 
MTPC analysis, important cuts in both analysis include the distance of closest approach between $\Lambda$ decay products, the distance of $\Xi$ and $\Lambda$ vertices from the interaction vertex, and the impact parameters of $\Xi$ decay products.
Fig.~\ref{Fig1} shows the invariant-mass distribution for $\Xi^-$ and $\overline{\Xi}^+$, reconstructed in the VTPC2, before background subtraction.
A mass resolution of FWHM=9 MeV for $\Xi^-$ and FWHM=11 MeV for $\overline{\Xi}^+$ was obtained. The invariant mass spectrum of $\Xi^-$+$\overline{\Xi}^+$ analyzed in the MTPC is broader (FWHM$\approx$30 MeV) due to the lack of direct momentum measurement
. The background was found to be combinatoric and, depending on cuts, in the range of a few percent in VTPC2, and about 20\% in MTPC. It was estimated by mixing $\Lambda$s ($\overline{\Lambda}$s) reconstructed in one event with all negative (positive) tra
cks from another event. The mixed events, processed in the same way as the real ones, resulted in a reconstructed ``combinatoric'' $\Xi$ ($\overline{\Xi}^+$) signal (and, in ``combinatoric'' $\Xi$+$\overline{\Xi}$ signal in case of MTPC), which was used, 
subsequently, for the evaluation of the magnitude of background corrections.
The reconstruction efficiency was estimated by embedding \cite{Mil} simulated decays generated by a GEANT \cite{CERN} based Monte Carlo into raw data events. The events with embedded cascades were treated analogously to the real data (pattern recognition,
 track fitting, hyperon selection). The efficiency was found to depend on rapidity and $p_{T}$; therefore each bin of ($y,p_{T}$) presented here has been corrected individually. The average overall tracking and reconstruction efficiency convoluted with th
e phase space acceptance was $\sim$0.8\% for the VTPC2 and $\sim$1\% for the MTPC.
Fig.~\ref{Fig2} shows transverse-mass spectra for the $\Xi^-$ and $\overline{\Xi}^+$ analyzed in the VTPC2, and $\Xi^-$ + $\overline{\Xi}^+$ analyzed in the MTPC.
The shape of the distributions is approximately exponential over the full range. For comparison (open circles) we also present the $\Xi^-$ + $\overline{\Xi}^+$ sum from the VTPC2 data sample. Both $\Xi^-$ + $\overline{\Xi}^+$ spectra are in agreement. The
 distribution is fitted with an exponential function in $m_{T}$:
\[
\frac{d^{2}n}{dm_{T}dy}=C(T)m_{T}e^{-\frac{m_{T}}{T}}
\]
with T as fit parameter. The normalization constant $C(T)$ is constrained to the experimental yield. The inverse-slope parameters for $\Xi$ particles in each data set are similar and in the vicinity of 300 MeV (see Fig.~\ref{Fig3}a,b) . This agrees well w
ith the dependence of the inverse slope T on the particle mass in Pb+Pb collisions established by NA49 measurements \cite{Bor}. A similar dependence is reported by the NA44 collaboration \cite{NA44}. The slope parameters increase monotonically with partic
le mass (in the case of NA49: from pions to deuterons), suggesting the presence of collective transverse flow in nuclear collisions at SPS energies \cite{App}. Our measurements of $\Xi$ inverse slope parameters are consistent with the slopes reported by t
he WA97 experiment \cite{And} for a more relaxed trigger.

The rapidity density distribution was obtained by integrating the transverse-mass spectra in two rapidity bins. They were chosen symmetrically with respect to $y_{mid}$=2.9. In the MTPC - the first bin covered $y\in$(1.7-2.2), the second - $y\in$(2.2-2.7)
; in the VTPC2 - the first bin covered $y\in$(3.1-3.6), the second - $y\in$(3.6-4.1).
The inverse slope parameters in both rapidity bins are rather similar as illustrated in Fig.~\ref{Fig3}a,b. The left panel shows inverse slope parameters for $\Xi^-$ + $\overline{\Xi}^+$ from the MTPC and VTPC2 and the right one for $\Xi^-$ from the VTPC2
 (the analogous plot for $\overline{\Xi}^+$ lacks in sufficient statistics). Full symbols are measured values, open symbols have been reflected at mid-rapidity. The transverse mass distribution was extrapolated to regions outside the NA49 acceptance by us
ing the parametrization given above.
In Fig.~\ref{Fig3}c,d the rapidity distributions for $\Xi^-$ + $\overline{\Xi}^+$ from the MTPC and VTPC2 (left) and for $\Xi^-$ from the VTPC2 (right) are presented. Full and open symbols represent measured and reflected values, as before. The two indepe
ndent measurements of $\Xi$ particles in the MTPC and VTPC2 agree within errors. At mid-rapidity, the $\Xi^-$ + $\overline{\Xi}^+$ rapidity density is measured to be about 2.7, with a major fraction of this number ($\sim$2.3) attributed to $\Xi^-$ rapidit
y density (right panel). The former value appears to be $\sim$30\% higher than the WA97 measurement \cite{And}. 
The $\overline{\Xi}^+$/$\Xi^-$ ratio at mid-rapidity is 0.23$\pm$0.03. This agrees within errors with the result of WA97 for somewhat less central Pb+Pb collisions \cite{Kra}. 
The systematic errors were studied in both detectors. The stability of the dependence of final results on the cuts applied showed that the systematic uncertainty in the integrated yields in both cases did not exceed 15\%. This is confirmed by the differen
ces in the $\Xi^-$ + $\overline{\Xi}^+$ mesurements in the MTPC and VTPC2 presented on Fig.~\ref{Fig3}c. 
The mid-rapidity multiplicities per unit of rapidity, dN/dy, are 2.29$\pm$0.12 and 0.52$\pm$0.05 for $\Xi^-$ and $\overline{\Xi}^+$, respectively. Quoted errors are statistical only.

 Making use of reflection symmetry we have systematically investigated \cite{Gab1} Gaussian extrapolations of the rapidity distributions in Fig.~\ref{Fig3}c,d. The width was assumed to be 1.1. This was motivated by the $\Xi$ rapidity distribution provided
 by the UrQMD model \cite{Bas} and the vector meson $\phi$ width measured in our experiment ($\sigma_{\scriptscriptstyle \phi}$=1.07) \cite{Fri}. The resulting estimates of total 4$\pi$ multiplicities are 8.2$\pm$1.1 for $\Xi^-$ + $\overline{\Xi}^+$ and 7
.4$\pm$1.0 for $\Xi^-$. These values are not corrected for the hyperon $\Omega^-$ feed-down: $\Omega^- \rightarrow \Xi^- +\pi ^0$ (B.R. = 8.3\%); however, based on the $\Omega$ yields in Pb+Pb collisions reported by WA97 \cite{And}, we estimate that this 
contribution is very small (less than 0.5\%). 
The systematic error of the extrapolation procedure, not included in the errors quoted above, was estimated to be less that 14\% by comparing the results obtained using different assumptions about the shape of the rapidity distribution \cite{Gab1}. 
As our central trigger (5\% of the total inelastic cross section) corresponds to a mean number of 370 participating nucleons we thus obtain an estimate of 0.020$\pm$0.002 $\Xi^-$s per participating nucleon and thus 0.040$\pm$0.004 per NN participant pair.
 In order now to check the degree of strangeness enhancement implied by these results we require similar data for elementary pp collisions. To our knowledge no such data exist in the vicinity of our $\sqrt{s}$=18 GeV but at $\sqrt{s}$=6\cite{Alp} and 63 G
eV\cite{Ake}. The $\Xi^-$/$\Lambda$ ratio is 0.016$\pm$0.011 and about 0.06$\pm$0.02\footnote{At the ISR energy ($\sqrt{s}$=63 GeV) we used the measured $\overline{\Xi}^+$/$\overline{\Lambda}$ ratio \cite{Ake} as an estimate of the $\Xi^-$/$\Lambda$ ratio
 at mid-rapidity. This provides an upper limit of the $\Xi^-$/$\Lambda$ ratio in 4$\pi$ acceptance.}, respectively for these two energies. 
Assuming that this ratio increases faster with $\sqrt{s}$ at the lower of the two energies we do not interpolate linearly but estimate the $\Xi^-$/$\Lambda$ ratio to be 0.03$\pm$0.01 at $\sqrt{s}$=18 GeV. Taking the $\Lambda$ multiplicity at this energy t
o be 0.10$\pm$0.03 from ref. \cite{Roh} we finally arrive at an estimate of a $\Xi^-$ multiplicity of 0.003$\pm$0.0015 in pp collisions at $\sqrt{s}$=18 GeV. The $\Xi^-$ yield per nucleon pair in central Pb+Pb collisions is thus enhanced by about one orde
r of magnitude, over and above the yield in elementary collisions. A more rigorous determination of the enhancement factor will be derived from analysis of pp collisions at $\sqrt{s}$=18 GeV currently in progress by NA49.

In summary, we have presented data on the absolute yields of $\Xi^-$ and $\overline{\Xi}^+$ as a function of rapidity near mid-rapidity in central Pb+Pb collisions at the SPS using two independent analysis techniques, one of which is a novel method incorp
orating topological identification without a magnetic field. We have reported a strong increase in the production of $\Xi^-$ and $\overline{\Xi}^+$ with respect to pp interactions. The inverse slope parameters of the transverse mass distribution of $\Xi$ 
particles are consistent with previously reported systematics \cite{Bor} reflecting the presence of transverse collective hadronic expansion flow.                                              
    
Acknowledgements: This work was supported by the Director, Office of Energy Rese
arch, 
Division of Nuclear Physics of the Office of High Energy and Nuclear Physics 
of the US Department of Energy under Contract DE-AC03-76SF00098, 
the US National Science Foundation, 
the Bundesministerium fur Bildung und Forschung, Germany, 
the Alexander von Humboldt Foundation, 
the UK Engineering and Physical Sciences Research Council, 
the Polish State Committee for Scientific Research (2 P03B 01912
and 2 P03B 9913), 
the Hungarian Scientific Research Foundation under contracts T14920 and T23790,
the EC Marie Curie Foundation,
and the Polish-German Foundation.

%
%
%
%
%
%
%
\end{document}